\documentclass[10pt]{article}
\usepackage{amsmath}
\usepackage{times}
\usepackage{ifpdf}
\usepackage[english]{babel}
\usepackage{hyperref}
\usepackage{mdwlist}

\usepackage{amsfonts}

\usepackage{amssymb} 

\usepackage{amsmath}
\usepackage{float}
\usepackage{graphicx}
\usepackage{graphics}
\usepackage[usenames]{color}
\usepackage{pdfpages}
\usepackage{vmargin}
\usepackage{subfigure}
\usepackage{hyperref}
\usepackage{ifthen}
\usepackage{theorem}
\usepackage{listings} 
\usepackage{amssymb} 
\usepackage[latin1]{inputenc}
\usepackage{amsmath}

\begin{document}
%



\title{Concurrent Constraint Conditional-Branching Timed Interactive Scores}

\author{Mauricio Toro$^{\rm a}$$^{\ast}$\thanks{$^\ast$Corresponding
    author. Email: mtoro [at] eafit.edu.co
\vspace{6pt}} , Myriam Desainte-Catherine$^{\rm b}$, \\\vspace{9pt} $^{\rm a}${\em{ Universidad EAFIT, Colombia}} \\ $^{\rm b}${\em{LaBRI,  
    Universit\'{e} de
   Bordeaux, France} } } 

\maketitle

\begin{abstract}
Multimedia scenarios have multimedia content and interactive events associated with computer programs.
Interactive Scores (IS) is a formalism to represent such scenarios by temporal objects, temporal relations (TRs) and interactive events.
IS describe TRs, but IS cannot represent TRs together with conditional branching.
We propose a model for conditional branching timed IS in the Non-deterministic Timed Concurrent Constraint (ntcc) calculus. We ran a prototype of our model in Ntccrt (a real-time capable interpreter for ntcc) and the response time was acceptable for real-time interaction.
An advantage of ntcc over Max/MSP or Petri Nets is that conditions and global constraints are represented declaratively.

\end{abstract}

\section{Introduction}\label{sec:introduction}
Interactive multimedia deals with the design of scenarios where multimedia content and interactive events can be associated with computer programs.
Designers usually create multimedia for their scenarios, then they bind them to external interactive events or programs. \textit{Max/MSP} and \textit{Pure Data (Pd)} \cite{max} are often used to program interactive scenarios. 
However, we claim for the need of a general model to (i) control synthesis based on human gestures and to (ii) declare relations among multimedia objects (e.g., partial-order relations for their execution).

\textit{Interactive Scores (IS)} is a formalism for the design of scenarios represented by \textit{temporal objects (TOs)}, \textit{temporal relations (TRs)} and interactive events. Examples of TOs are videos and sounds. TOs can be triggered by interactive events (usually launched by the user) and several TOs can be active simultaneously. A TO can contain other TOs. The hierarchy allows us to control the start or end of a TO by controlling the start or end of its parent. Moreover, TRs provide a partial order for the execution of the TOs: TRs can be used to express precedence between objects.


IS have been subject of study since the beginning of the century \cite{bdc01}, \cite{bdc03}. IS were originally developed for interactive music scores.
Recently, the model was extended by Allombert, Desainte-Catherine, Larralde and Assayag in \cite{artech2008}.
Hence IS can describe any kind of TOs, Allombert \textit{et al.}'s model has inspired two applications: \textit{iScore} \cite{aadc08} to compose and perform Electroacoustic music and \textit{Virage} \cite{virage} to control live spectacles and interactive museums.

IS are successful to describe TRs, but IS have not been used to represent TRs together with \textit{conditional branching}.
Conditional branching is used in programming to describe control structures such as \textit{if/else} and \textit{switch/case}.
It provides a mechanism to choose the state of a program depending on a condition and its current state.

Using conditional branching, a designer can create scenarios with loops and choices (as in programming).
The user and the system can take decisions during performance 
with the degree of freedom described by the designer --while the system maintains the TRs of the scenario.

The designer can express under which conditions a loop ends; for instance, when the user changes the value of a certain variable, the loop stops; or the system non-deter- ministically chooses to stop. 

Unfortunately, there is neither a theoretical model nor a special-purpose application to support conditional branching in interactive multimedia.
In this work, we propose a model for conditional-branching timed IS in the \textit{Non-deterministic Timed Concurrent Constraint} (\texttt{ntcc}) \cite{ntcc} calculus. In our model we combine TRs, conditional branching and discrete interactive events in a single model. We ran a prototype of the model over \textit{Ntccrt} \cite{ntccrt}, a real-time capable interpreter for \texttt{ntcc}.

In a previous work \cite{tbdcb10}, we showed how we can represent a multimedia installation with loops and choice\footnote{\url{http://www.gmea.net/activite/creation/2007_2008/pPerez.htm}}, and the pure timed IS model \cite{artech2008} into our model.

\subsection{Related work on interactive multimedia}
A similar approach to ours was followed by Olarte and Rueda in \cite{or09a}. They propose a model for IS in a calculus similar to \texttt{ntcc}; however, they only modeled TRs. They verified critical properties on the system. The key point of their model is that the user can change the hierarchical structure of the score during performance.

Another system dealing with a hierarchical structure is \textit{Maquettes of OpenMusic} \cite{Bresson05b}. However,
OpenMusic is a software for composition and not real-time interaction.

Another kind of systems capable of real-time interaction are \textit{score following} systems (see \cite{cont08a}).
Such systems track the performance of a real instrument and they may play multimedia associated to certain notes of the piece.
However, to use these systems it is necessary to play a real instrument; whereas to use IS, the user only has to control some parameters of the piece, such as the start and end dates of the TOs.

A model for multimedia interaction that does not require a real instrument uses Hidden Markov Models to model probabilistic installations \cite{bsf07}. The system tracks human motion and it responds to human performance with chords and pitches depending on the knowledge of 
previous training. However, the system requires intensive training and it is not a tool for composition.

In the domain of composition of interactive music, there are applications such as \textit{Ableton Live}\footnote{\url{http://www.ableton.com/live/}}. Using \textit{Live}, a composer can write loops and a musician can control
different parameters of the piece during performance. Live is commonly used for Electronic and Electroacoustic music. Unfortunately, the means of interaction and the synchronization patterns provided by Live are limited.

\subsubsection{Formalisms for Interactive Multimedia}
 To handle complex synchronization patterns and to predict the behavior of interactive scenarios, 
 formalisms such as \texttt{ntcc} and \textit{Hierarchical Time Stream Petri Networks (HTSPN)} \cite{SSW95} and have been used to model IS \cite{AADR06, artech2008}.

 In HTSPN we can express a variety of TRs, but it is not easy to represent global constraints (e.g.,
  the number of TOs playing simultaneously). Instead, \texttt{ntcc} synchronizes processes through a common 
 \textit{constraint store}, thus global constraints are explicitly represented in such store. We chose \texttt{ntcc} because we can easily represent time, constraints, choice, and we can verify the model.

Another formalism for defining declaratively partial orders of musical processes and audio is Tempo \cite{tempo}. 
However, Tempo does not allow us to express choice (when multiple conditions hold), simultaneity and weak time-outs (e.g., perform an action if the condition cannot be deduced). A key aspect is that there is a real-time capable interpreter and
automatic verification for Tempo. 

At present, there is not an automatic verifier for \texttt{ntcc}.
In the declarative view, \texttt{ntcc} processes can be interpreted as \textit{linear temporal logic} formulae. \texttt{Ntcc} includes an inference system in this logic to verify properties of  \texttt{ntcc} models. This inference procedure was proved to be of exponential time complexity \cite{ntcc-phd}. Nevertheless, we believe practical automatic verification could be envisioned  for useful subsets of \texttt{ntcc} via model checking (see \cite{fv06}). 

Automated verification for IS will provide information about the correctness of the system to computer scientists. It will also provide important properties about the scenario to its designers and users. 
It will be possible to verify the absence of deadlocks, and also that certain TOs will be played during performance. This kind of properties cannot be verified in applications with no formal semantics.

\subsection{Structure of the paper}
The remainder of this paper is structured as follows. Section 2 explains \texttt{ntcc} and Ntccrt. Section 3 states our model for conditional-branching timed IS. Section 4 shows the \texttt{ntcc} definitions of our model. Section 5 explains our implementation using Pd and Ntccrt. Finally, section 6 gives some concluding remarks and future work.

\section{The ntcc process calculus}\label{sec:ntcc}
A family of process calculi is \textit{Concurrent Constraint Programming} (\texttt{ccp}) \cite{ccp}, where a system is modeled in terms of variables and constraints over some variables. The constraints are contained in a common \textit{store}. There are also agents that reason about the system variables, based on  partial information (by the means of constraints).

Formally, \texttt{ccp} is based upon the idea of a \textit{constraint system (CS)}. A constraint system includes a set of (basic) constraints and a relation (i.e., entailment relation $\models$) to deduce a constraint with the information supplied by other constraints.

A \texttt{ccp} system usually includes several CSs for different variable types. 
 There are CSs for variable types such as sets, trees, graphs and natural numbers. A CS providing arithmetic relations over natural numbers is known as \textit{Finite Domain (FD)}. As an example, using a FD CS, we can deduce $pitch \neq 60$ from the constraints $pitch > 40$ and $pitch < 59$.

Although we can choose an appropriate CS to model any problem, 
in \texttt{ccp} it is not possible to delete nor change information accumulated in the store. For that reason it is difficult to perceive a notion of discrete time, useful to model reactive systems communicating with an external environment (e.g., users, lights, sensors and speakers).

\texttt{Ntcc} introduces to \texttt{ccp} the notion of discrete time as a sequence of \textit{time units}. Each time unit starts with a store  (possibly empty) supplied by the environment, and \texttt{ntcc} executes all the processes scheduled for that time unit.
In contrast to \texttt{ccp}, in \texttt{ntcc} we can model variables changing values over time. 
A variable $x$ can take different values at each time unit. To model that
in \texttt{ccp}, we have to create a new variable $x_i$ each time we change the value of $x$.


Process calculi has been applied to the modeling of interactive music systems
 \cite{toro2016faust, toro2016gelisp, 2016arXiv160202169T, toro2015ntccrt, is-chapter,tdcr14,ntccrt,cc-chapter,torophd,torobsc,Toro-Bermudez10,Toro15,ArandaAOPRTV09,tdcc12,toro-report09,tdc10,tdcb10,tororeport} 
 and ecological systems \cite{EPTCS2047, PT13,TPSK14,PTA13,mean-field-techreport}. 
\subsection{Ntcc in multimedia interaction}

In this section we give some examples on how the computational agents of \texttt{ntcc} can be used with a FD CS.
A summary of the agents semantics can be found in Table \ref{tab:ntccagents}.

\begin{table}[!h]
  \begin{center}   
     \begin{tabular}{|ll|}
\hline
Agent & Meaning\\
\hline
\textbf{tell} $(c)$ & Adds $c$ to the current store\\
\textbf{when} $(c)$ \textbf{do} $A$  & If $c$ holds now run $A$\\
\textbf{local} $(x)$ \textbf{in} $P$  & Runs $P$ with local variable $x$\\
$A$ $\|$ $B$ & Parallel composition \\
\textbf{next} $A$ & Runs  $A$ at the next time-unit \\
\textbf{unless} $(c)$ \textbf{next} $A$  & Unless $c$ holds, next run $A$ \\
$\sum _{i \in I}$ \textbf{when} $(c_{i})$ \textbf{do} $P_{i}$  & Chooses  $P_{i}$ s.t. $(c_{i})$ holds \\ 
*$P$ & Delays P indefinitely \\ 
!$P$ & Executes P each time-unit\\
\hline
\end{tabular}    
    \caption{Semantics of \texttt{ntcc} agents.}
    \label{tab:ntccagents}
  \end{center}
\end{table}

\begin{itemize}
\item Using \textit{tell} it is possible to add constraints to the store such as $\textbf{tell} (60 < pitch_2 < 100)$, which means that $pitch_2$  is an integer between 60 and 100. 

\item \textit{When} can be used to describe how the system reacts to different events; for instance, 
$\textbf{when}$ $pitch_1=C4 \wedge pitch_2 = E4 \wedge pitch_3 = G4$ $\textbf{do}$   $\textbf{tell} (CMayor = true)$ adds the constraint 
$CMayor = true$ to the current store as soon as the pitch sequence C, E, G  has been played.

\item \textit{Parallel composition} ($\|$) makes it possible to represent concurrent processes; for instance,  
$\textbf{tell}$ $(pitch_1 = 52)$ $\|$ $\textbf{when}$ $48 < pitch_1 < 59$ $\textbf{do}$ $\textbf{tell}$ $(Instrument = 1)$ 
tells the store that $pitch_1$ is 52 and concurrently assigns the \textit{instrument} to one, since $pitch_1$ is in the desired interval (see fig. \ref{fig:tellwhenpar}).

\begin{figure}[!h]
 \centerline{\framebox{
 \includegraphics[width=\columnwidth]{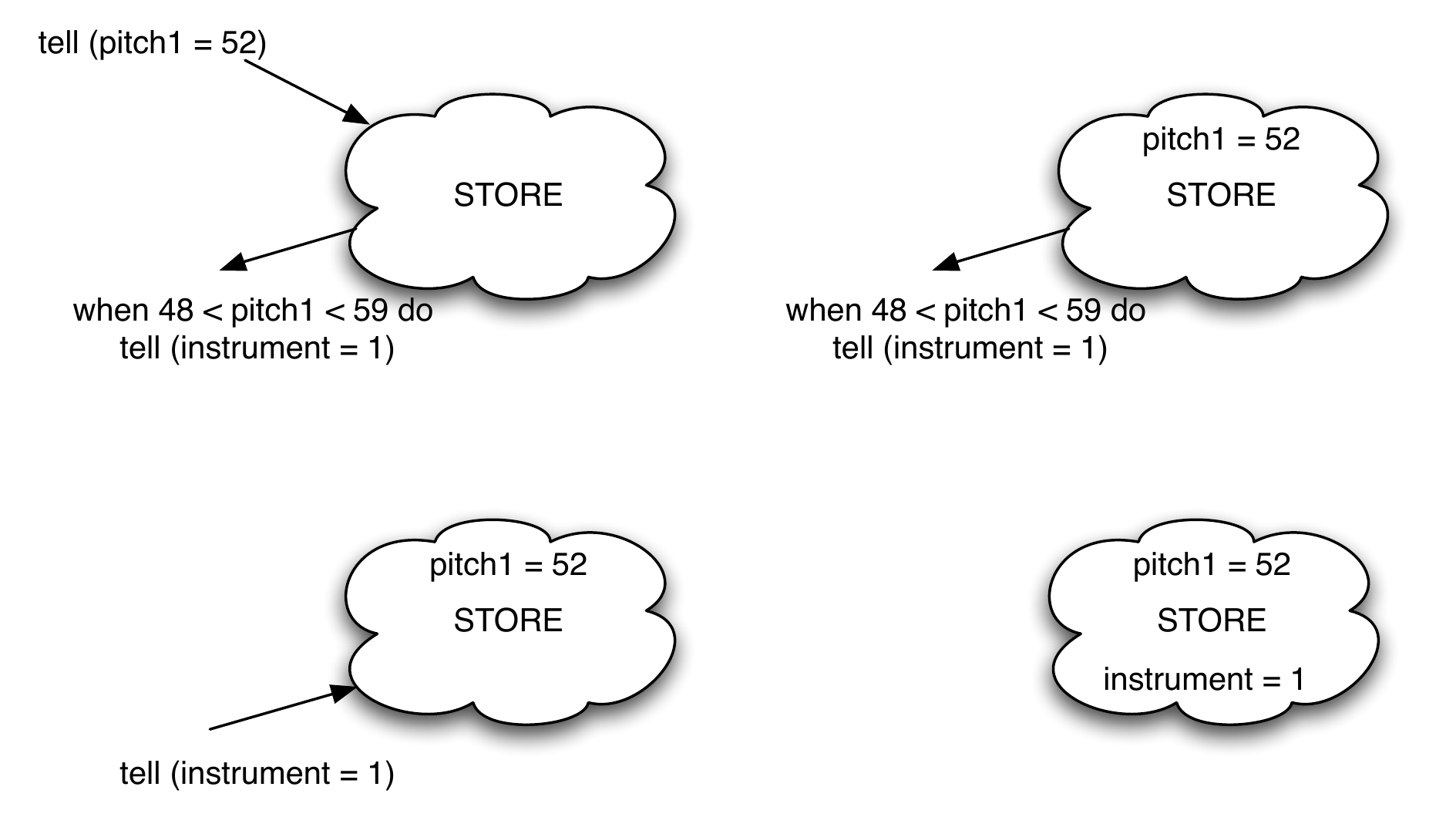}
}}
 \caption{An example of the \texttt{ntcc} agents.}
 \label{fig:tellwhenpar}
\end{figure}

\item \textit{Next} is useful
when we want to model variables changing over time; for instance, $ \textbf{when}$ $(pitch_1 = 60)$  $\textbf{do}$ $\textbf{next}$  $\textbf{tell}$ $(pitch_1 <> 60)$ means that if $pitch_1$ is equal to 60 in the current time unit,  it will be different from 60 in the next time unit.

\item \textit{Unless} is useful to model systems reacting when a condition is not satisfied or when the condition cannot be deduced from
the store; for instance, $\textbf{unless}$ $(pitch_1 = 60)$  $\textbf{next}$ $\textbf{tell}$ $(lastPitch <> 60)$ reacts when $pitch_1 = 60$ is false or when $pitch_1 = 60$ cannot be deduced from the store (e.g., $pitch_1$ was not played in the current time unit).

\item \textit{Star} (\textbf{*}) can be used to delay the end of a process indefinitely, but not forever; for instance, $* \textbf{tell}$ $(End = true)$. Note that to model Interactive Scores we do not use the \textit{star} agent.

\item \textit{Bang} ($\textbf{!}$)  executes a certain process every time unit after its execution; for instance, $!$$\textbf{tell}$ $(C_4 = 60)$. 

\item \textit{Sum} (\textbf{$\sum$}) is used to model non-deterministic choices; for instance, $\sum_{i \in \{48,52,55\}}$ \textbf{when} $i \in PlayedPitches$
\textbf{do} $\textbf{tell}$ $(pitch = i)$ chooses a note among those played previously that belongs to the C major chord. 

\end{itemize}

In \texttt{ntcc}, recursion can be defined (see \cite{ntcc-phd}) with the form $q(x) =^{def} P_q$, where $q$ is the process name and
$P_q$ is restricted to call $q$ at most once and such call must be within the scope of a \textit{next}.
The reason of using \textit{next} is that \texttt{ntcc} does not allow 
recursion within a time unit.

The reader should not confuse a simple definition with a recursive definition; for instance, $Before_{i,j}$ $=^{def} \textbf{tell} (i \in Predecessor_j)$ is a simple
definition where the values of $i$ and $j$ are replaced statically, like a macro in a programming language.
Instead, a recursive definition such as $Clock(v)$ $=^{def} \textbf{tell} (clock=v) \| \textbf{next}\ Clock(v+1) $
is like 
a function in a programming language.

\subsection{Ntccrt: A real-time capable interpreter for ntcc}
In the current version of Ntccrt, we can write a \texttt{ntcc}
model on either Lisp, Openmusic or C++. For a complete implementation of Interactive Scores, it will be necessary to produce automatically the corresponding \texttt{ntcc} model based on a graphical interface similar to Virage. 

\begin{figure}[!h]
 \centerline{\framebox{
 \includegraphics[width=\columnwidth]{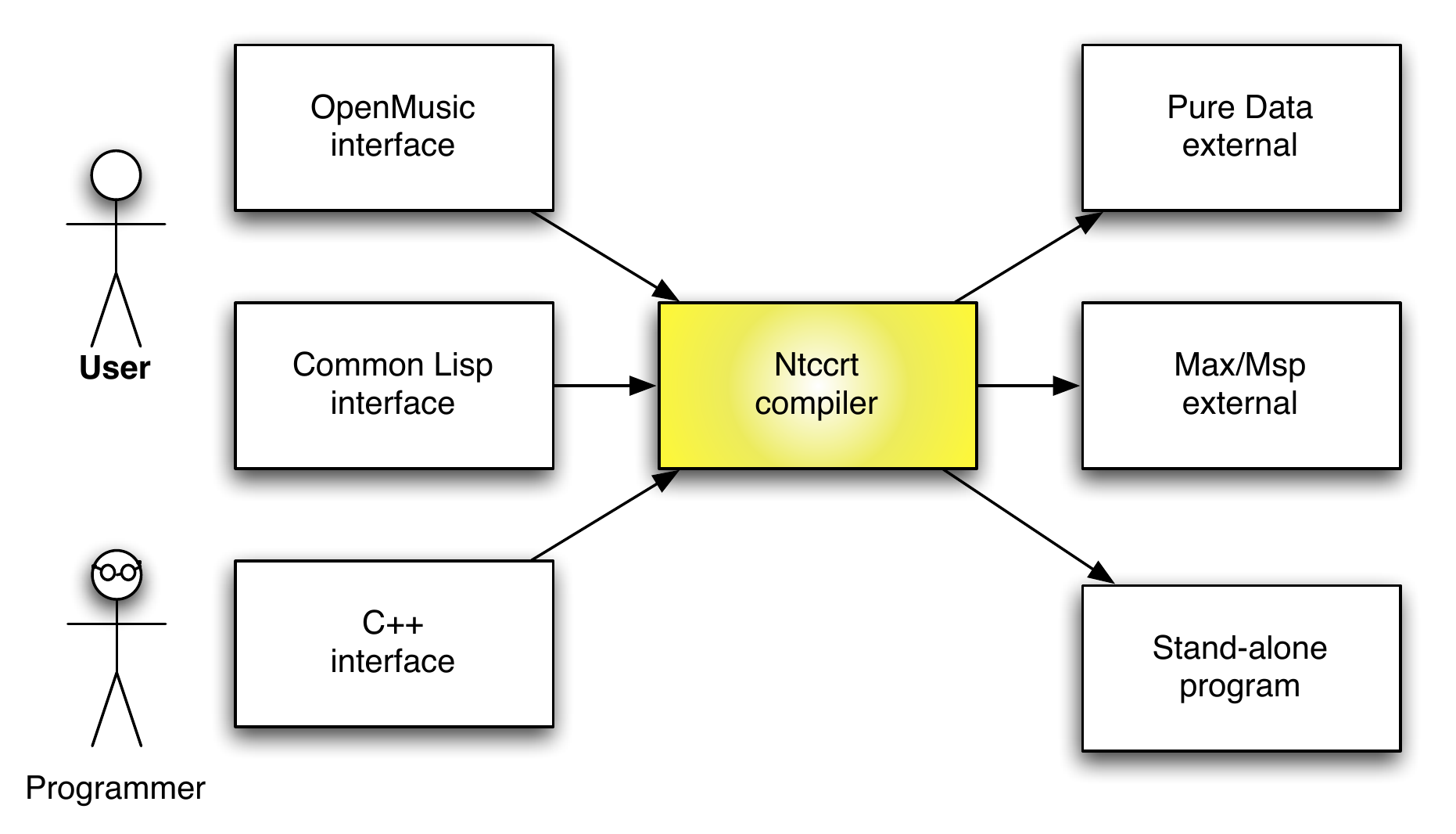}
}}
 \caption{Interfaces of Ntccrt.}
 \label{fig:ntccrt}
\end{figure}

To execute a \texttt{ntcc} model it is not necessary to develop an interface because Ntccrt programs can be compiled into stand-alone programs or as
external objects (i.e., a binary plugins) for Pd or Max (see fig. \ref{fig:ntccrt}).

We can use the message passing API provided by Pd and Max to communicate any object with the Ntccrt external. We can also control all the available objects for audio and video processing defined in those languages using Ntccrt.
To synchronize those objects, Ntccrt provides an important part of Gecode's constraints  \cite{gecode}.

Ntccrt uses Gecode as its constraint solving library. Contraint solving libraries can be used to solve combinatory problems such as
planning systems optimal delivery of packages \cite{PAT2016}; however, they
can also be used for constriant propagation. Gecode was carefully designed to
support efficiently the Finite Domain (FD) constraint system. Ntccrt relies on propagation of FD constraints. 

\section{Conditional branching timed IS}\label{sec:typeset_text}


Points and intervals build up Interactive Scores (IS), thus  
a \textit{score}\footnote{We still use the term \textit{score} for historical reasons.} (i.e., the specification of a scenario) is defined by a tuple $s = \langle P,I \rangle$, where $P$ is a set of points and $I$ is a set of intervals. A temporal object is just a type of interval.  

\subsection{Points}
Intuitively, a point $p$ is a \textit{predecessor} of $q$ if there is a relation $p$ \textit{before} $q$. Analogically,
a point $p$ is a \textit{successor} of $r$ if there is a relation $r$ \textit{before} $p$.

A \textit{Point} is defined by $p = \langle  b_p, b_s \rangle$, where $b_p$ and $b_s$ represent the behavior of the point. Behavior $b_p$ defines
whether the point waits until all its predecessors transfer the control to it --\textit{Wait for All (WA)}-- or it only waits for the first of them --\textit{Wait for the First (WF)}--. Behavior $b_s$ defines whether the point transfers the control to all its successors which conditions hold --\textit{No CHoice (NCH)}--  or it chooses one of them --\textit{CHoice (CH)}--.

Note that we do not include the set of dates of the point in previous definition. Beuriv\'{e} \textit{et al.} argued in \cite{bdc01} that the edition of a hierarchical representation of music using a relative time model requires less variable updates than using an absolute time model.
We argue that it is also true during the performance of Interactive Scores. Moreover, in our model it is not easy to know
the set of all possible dates \textit{a priori} because they depend on the choices that the user makes during performance.




\subsection{Intervals: TCRs and TOs}
An interval $p$ \textit{before} $q$ intuitively means that
the system waits a certain time to transfer the control from $p$ to $q$ if the condition in the interval holds. In addition, it executes a process  throughout its duration.
An interval also
has a \textit{nominal duration} that may change during the performance. The nominal duration is computed during the edition of the scenario using constraint programming (see \cite{AADR06}. Formally, an interval is a tuple composed by 
\begin{itemize*}
\item a start point $(p_1)$
\item an end point $(p_2)$
\item a condition $(c)$
\item a duration $(d)$
\item an interpretation for the condition $(b)$
\item a local constraint $(l)$
\item a process $(proc)$
\item parameters for the process $(param)$
\item children $(N)$
\item local variables $(vars)$
\end{itemize*}


It is not practical to include all those elements explicitly; thus, we have identified two types of intervals.
\textit{timed conditional relations (TCRs)} have a condition $c$ and an interpretation $b$, but they do not have children, their local constraint is \texttt{true}, and their process
is $silence$\footnote{$silence$ is a process that does nothing.}. \textit{Temporal objects (TOs)} may have children, local variables and a local constraint, but their condition is \texttt{true}, and their interpretation is $when$ (i.e., when the condition is true, it transfers the control from $p_1$ to $p_2$).

To have a coherent score, we must define a TCR between the start point of each father and the start point of at least one of its children. However, it is not required to connect a child to the end point of its father.
Furthermore, in our model we may define multiple TCRs and TOs between two points. This does not introduce an incoherence in the model because the behavior of those intervals (as any interval) depends on the behavior of the points and the parameters of the interval.

\subsubsection{Timed Conditional Relations (TCRs)}
A \textit{timed conditional relation (TCR)} is defined by $r = \langle p_1,\\ p_2, c, d, b\rangle$, where $p_1$ and $p_2$ are the points involved in the relation. The condition $c$ determines whether the control \textit{jumps} from $p_1$ to $p_2$ (i.e., the control is transferred from $p_1$ to $p_2$). The interpretation of $c$ is $b$. 
 There are two possible values for $b$: (i)
\textit{when} means that if $c$ holds, the control jumps to $p_2$; and (ii) \textit{unless} means that if $c$ does not hold or its value cannot be deduced from the environment (e.g., $c = a > 0$ and $-\infty < a < \infty$), the control jumps to $p_2$.  

A duration is \textit{flexible} if it can take any value, \textit{rigid} if it takes values between two fixed integers and \textit{semi-rigid} if it takes values greater than a fixed integer. 
In our model, we always respect flexible durations. 
Our model is based upon transferring the control from one point to another. For that reason, it is not always
possible to respect rigid and semirigid durations; for instance, when a point waits for an event or when 
it is followed by a choice.



\subsubsection{Temporal objects (TOs)}
A \textit{temporal object (TO)} is defined by $t = \langle p_s, p_e, l, d, proc, \\param, N , vars \rangle$ where $p_s$ is a point that starts a new instance of $t$ and $p_e$ ends such instance.  
A constraint $l$ is attached to $t$, it contains local information for $t$ and its children. The duration is $d$. A process which executes throughout the duration of $t$ is $proc$.  The list of parameters for the process is $param$.  
The set of TOs embedded in $t$ is $N$, which are called children of $t$. 
 Finally, $vars$ represents the local variables defined for the TO that can be used by $t$'s children, process and local constraint.


\subsection{Example: A loop controlled by a condition}\label{subsection:example}
\begin{figure}[h!]
 \centerline{\framebox{
 \includegraphics[width=\columnwidth]{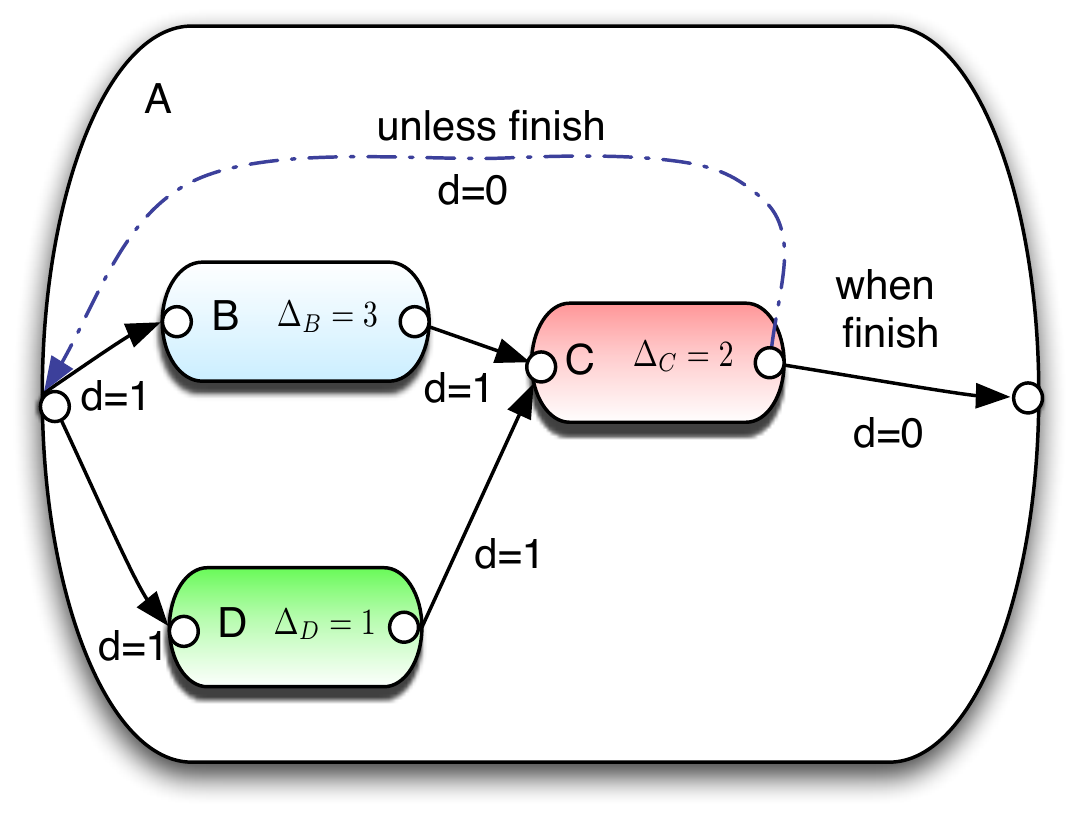}
}}
 \caption{A score with a user-controlled loop.}
 \label{fig:basic-example}
\end{figure}

The following example (see fig. \ref{fig:basic-example}) describes a score with a loop. During the execution, the system plays a silence of one second. After the silence, it plays the sound $B$ during three seconds and simultaneously it turns on the lights $D$ for one second. After the sound $B$, it plays a silence of one second, 
then it plays video $C$. If the variable $finish$ becomes true, it ends the scenario after playing the video $C$; otherwise, it
jumps back to the beginning of the first silence after playing the video $C$. 

To define the score of this scenario, we define a local boolean variable
$finish$ in $A$, and we use it as the condition for some TCRs. 
Note that the silence between $D$ and $C$ lasts one second in the score, but during execution it is longer because of the behavior of the points.

The points have the following behavior. The end point of $C$ ($e_c$) is enabled for choice, and the other points transfer the control to all their successors. The start point of $C$ ($s_c$) waits for all its predecessors to transfer the control to it, and all the other points wait for the first
predecessor that transfers the control to them. 

Formally, the points are defined\\

\noindent
$s_a = e_a = s_b = e_b = s_d = e_d =\langle\{ WF, NCH\}\rangle$\\
$s_c = \langle\{ WA, NCH\}\rangle$\\
$e_c = \langle\{ WF, CH\}\rangle$\\
$P = \{s_a,e_a,s_b,e_b,s_c,e_c,s_d,e_d\}$ \\

\noindent
As an example, $e_c$ Waits for the first predecessor (WF) and makes a choice (CH). 

The TOs are defined by\\

\noindent
$ A=\langle s_a,e_a, d \in [0, \infty), d , sil., \emptyset, \{B,C,D\}, \{finish\} \rangle\\
B=\langle s_b, e_b, \texttt{true}, 3, playSoundB, \emptyset, \emptyset, \emptyset\rangle\\
C=\langle s_c, e_c, \texttt{true}, 2, PlayVideoC, \emptyset, \emptyset, \emptyset\rangle\\
D=\langle s_d, e_d, \texttt{true}, 1, TurnOnLightsD, \emptyset, \emptyset, \emptyset\rangle \\
T = \{A,B,C,D\}$ \\

\noindent
As an example, $A$ is composed by points $s_a$ and $e_a$, it has a flexible duration, its process is silence, its children are $B$, $C$ and $D$ and its local variable is $finish$. 

In what follows we present the TCRs \\

\noindent
$TCR =\\
\hspace*{8pt}\{ \langle s_a,s_b,\texttt{true},1, when\rangle,
\langle s_a,s_d, \texttt{true}, 1, when\rangle,\\
\hspace*{12pt}\langle e_b,s_c, \texttt{true}, 1, when\rangle,
\langle e_d,s_c, \texttt{true}, 1, when\rangle,\\
\hspace*{12pt}\langle e_c,s_a, \neg finish, 0, when\rangle,
\langle e_c, e_a, finish, 0, when\rangle \}$\\ 

\noindent
As an example, the first one is a TCR between points $s_a$ and $s_b$, its condition is \texttt{true}, its interpretation is $when$ and its duration is one. 

Finally, $I$ is the set of intervals composed by the TOs and the TCRs and $S$ is the score. \\

\noindent
$I = T \bigcup TCR$\ \ \ \ \ \ \ \ \ $S = \{P,I\}$

\subsection{Limitations: Rigid durations and choice}

In some cases (e.g., fig. \ref{fig:basic-example}), we can respect rigid durations of TOs during performance. 
Unfortunately, there is not a generic way to compute the value of a rigid duration in a score with conditional branching. 
The problem is that choices do not allow us to predict the duration of a TO's successor; therefore, it is not possible to determinate \textit{a priori} the
duration of all the TOs. 

\begin{figure}[h!]
 \centerline{\framebox{     
\includegraphics[width=\columnwidth]{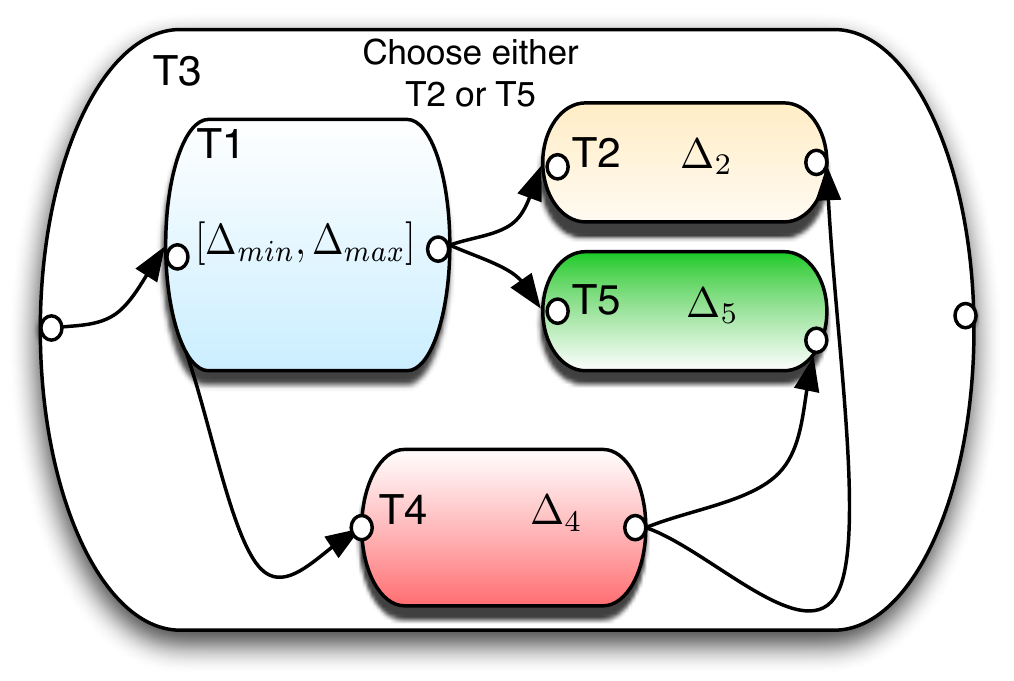}
      }}
    \caption{Limitation of rigid durations.}
    \label{fig:limitations}
\end{figure}

Figure \ref{fig:limitations} shows a scenario where we cannot respect rigid durations. $T_2$, $T_4$ and $T_5$ have fixed durations,
but $T_1$ can take different values between $\Delta_{min}$ and $\Delta_{max}$. Since there is no way to predict whether $T_2$ or $T_5$ will be chosen after the execution of $T_1$, we cannot compute a coherent duration for $T_1$ before the choice.


\section{Our ntcc model of IS}
In this section we define our \texttt{ntcc} model. We define processes for some combinations of the behaviors of a point.
The definition of an interval can be used for both timed conditional relations and temporal objects.
To represent intervals we create a graph with the predecessors and successors of each point using the variables $Predec$ and $Succ$. For simplicity, we do not include hierarchy, we only model the interpretation $when$, we can only declare a single interval between two points, and we can only execute a single instance
of an interval at the same time.

\subsection{Points: Three combinations of behaviors}
We only include three type of points: points that choose among their successors ($ChoicePoint$), points
that transfer the control to all their successors ($JumpToAllPoint$), and points that wait for all their predecessors to transfer the
control to them ($WaitForAllPoint$). The first two types of points wait for the first predecessor that transfers the control to them to be active.

Points are modeled using Finite Domain constraints; for instance, to know if at least one point has transferred the control to the  
point $i$, we ask to the store if the \textit{boolean or} ($\bigvee_{j \in P} $) constraint applied to the relation $Arrived(i,j)$ can be deduced from the store (where $P$ is the set of identifiers for each point).

When all the expected predecessors transfer the control to the point $i$, we say that the point is active (i.e., $ActivePoints_i$ holds).
Analogaly, when a point $i$ transfers the control to a point $j$, we add the constraint\\ $ControlTranferred(j,i)$.

In order to represent the choice between points $a$ and $b$, we use the variable $finish$ in the $\Sigma$ process. Note that $\textbf{when}\ c_1\ \textbf{do}\ P_1 + \textbf{when}\ c_2\ \textbf{do}\ P_2$ is equivalent to $\sum_{i \in \{1,2\}} \\ \textbf{when}\ c_i\ \textbf{do}\ P_i$, and $\textbf{whenever}\ c\ \textbf{do}\ P$ is equivalent to $\textbf{!when}\ c\ \textbf{do}\ P$.\\ 

\noindent
$ChoicePoint_{i,a,b}$ $\overset{def}{=}$ \\
\hspace*{07pt} \textbf{whenever} $\bigvee \limits_{j \in P} Arrived(i,j)$ \textbf{do} (\textbf{tell} $(ActivePoints_i)$ \\
\hspace*{12pt}$\|$ \textbf{when} $finish$ \textbf{do} \textbf{tell} ($ControlTransferred(a,i)$) \\
\hspace*{16pt}+\textbf{when}  $\neg finish$  \textbf{do} \textbf{tell} ($ControlTransferred(b,i)$)) \\

The following definition uses the agent $\prod$ to transfer the control to all the successors of the point $i$. The agent $\prod$ represents the parallel composition in a compact way. \\

\noindent
$ToAll_i$ $\overset{def}{=}$ \\
\hspace*{08pt}\textbf{tell} $(ActivePoints_i)$  \\
\hspace*{10pt}$\|$$\prod _{j \in P}$ \textbf{when} $Succs(i,j)$ \textbf{do} \\
\hspace*{60pt}\textbf{tell} ($ControlTransferred(j,i)$))\\

Using the definition $ToAll_i$, we define the two points that transfer the control to all its successors. \\

\noindent
$JumpToAllPoint_i$ $\overset{def}{=}$ \\
\hspace*{03pt} \textbf{whenever} $\bigvee _{j \in P} Arrived(i,j)$ \textbf{do} $ToAll_i$ \\

To wait for all the predecessors, we ask the store if the constraint $Arrived = Predec$ holds. \\

\noindent
$WaitForAllPoint_i$ $\overset{def}{=}$ \\
\hspace*{03pt} \textbf{whenever} $\forall j, Arrived(i,j) = Predec(i,j)$ \textbf{do} $ToAll_i$

\subsection{Intervals: TCRs and TOs}
Intervals are modeled by two recursive definitions. These definitions model both TOs and TCRs because intervals only change the value of an $ActivePoints$ variable, thus they only control the start and end of their processes. 

Process $I$ waits until at least one point transfers the
control to its start point $i$, and at least one point has been chosen by another point to transfer the control to its destination $j$.
When such conditions hold, it waits until the duration of the interval is over\footnote{\textit{next$^d$} is a process \textit{next} nested $d$ times ($next(next(next...$).}, then it transfers the control
from point $i$ to $j$. It also adds a constraint on the corresponding set of predecessors and successors. \\

\noindent
$I_{i,j,d}$ $\overset{def}{=}$ !(\textbf{tell} ($Predec(j,i)$) $\|$ \textbf{tell} ($Succ(i,j)$))\\
\hspace*{03pt} $\|$\textbf{whenever} $\bigvee _{k \in P} ControlTransferred(j,k)$ \\
\hspace*{45pt}$\land \bigvee _{k \in P} Arrived(i,k)$ \textbf{do}(\\ 
\hspace*{15pt}\textbf{next$^d$}(\textbf{tell}$(Arrived(j,i))$ $\| PredecessorsWait(i,j)$)) \\




$PredecessorsWait$ adds the constraint $Arrived(j,i)$ until the time unit after the point $j$ becomes active. This definition maintains the coherence of $WaitForAll$ points.\\

\noindent
$PredecessorsWait_{i,j}$ $\overset{def}{=}$ \textbf{unless} $ActivePoints_{j}$ \textbf{next} \\
\hspace*{37pt}$(PredecessorsWait_{i,j} \|$ \textbf{tell} ($Arrived(j,i)$)) 

\subsection{The example \ref{subsection:example} on ntcc}
The example presented on figure \ref{fig:basic-example} can be easily modeled in \texttt{ntcc}.
$User$ is a process representing a user that tells to the store that $finish$ is not true during the first $n$ time units, then it tells that $finish$ is true. Note that an advantage of \texttt{ntcc} is that the constraint
$i \geq n$ can be easily replaced by more complex ones; for instance, it can be replaced by 
$i \geq n \land c$. Constraint $c$ can be, for instance, ``there are only three active points at this moment in the score'' (i.e., $|\{ x \in ActivePoints\ |\ x = 1  \} |=3$). \\ 

\noindent
$User_n(i)$ $\overset{def}{=}$ \textbf{when} $i \geq n$ \textbf{do} \textbf{tell} $(finish)$ \\
\hspace*{50pt} $\|$\textbf{unless} $i \geq n$ \textbf{next} \textbf{tell} $(\neg finish)$ \\
\hspace*{50pt} $\|$\textbf{next} $User_n(i+1)$\\

\noindent
$TCRs$ $\overset{def}{=} I_{s_d,e_d,1} $ \\
\hspace*{03pt}$\| I_{s_a,s_b,1} \| I_{e_d,s_c,1} \|I_{s_b,e_b,3} \|I_{e_b,s_c,1} \|I_{s_c,e_c,2} \|I_{e_c,s_a,0} \\
\hspace*{03pt}\|I_{e_c,e_a,0} \|I_{null,s_a,0} \|I_{s_a,s_d,1} \|$ \textbf{tell} ($Arrived(s_a,start)$) \\

\noindent
$Points$ $\overset{def}{=}$ $ChoicePoint_{e_c,e_a,s_a} \| WaitForAllPoint_{s_c} \\
\hspace*{47pt}\| \prod_{i \in \{s_a,e_a,s_b,e_b,s_d,e_d\}} JumpToAllPoint_i$ \\

\noindent
$System_n$ $\overset{def}{=}$ $ User_n(0) \| TCRs \| Points $

\section{Implementation in Ntccrt and Pd}
\begin{figure}[!h]
 \centerline{\framebox{
 \includegraphics[width=\columnwidth]{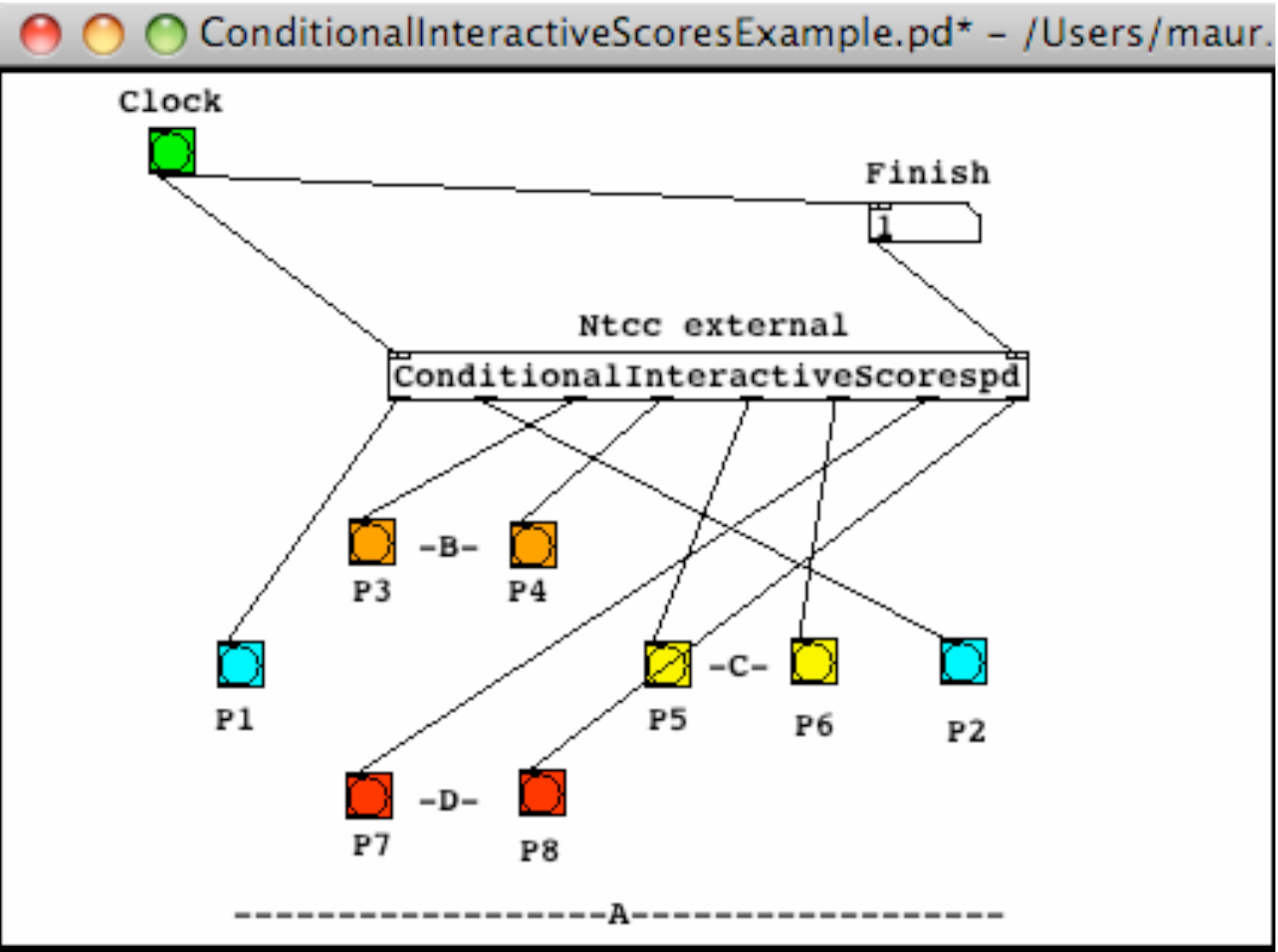}
}}
 \caption{Executing Example \ref{subsection:example} in Pd.}
 \label{fig:pd}
\end{figure}

We implemented the previous example in Ntccrt and Pure Data (Pd) (fig. \ref{fig:pd}).
We replaced the $User$ process with a user input for the variable $finish$. We generated a Ntccrt external (i.e., a binary plugin) for Pd with our \texttt{ntcc} model.

The  external has two inputs:
one for the clock ticks and one for the value of $finish$. 
The input for the clock ticks can be connected to a \textit{metronome} object to have a fixed duration for every time unit during the performance. The reader can find a discussion of executing time units with fixed durations in \cite{ntccrt}.

The Ntccrt external outputs a boolean value for each point, indicating whether it is active or not. Using such values, 
we can control the start and end of $SoundB$, $VideoC$ and $lightsD$, which are processes
defined in Pd.

\subsection{Results: Performance and usability of Ntccrt}
\begin{figure}[!h]
 \centerline{\framebox{
 \includegraphics[width=\columnwidth]{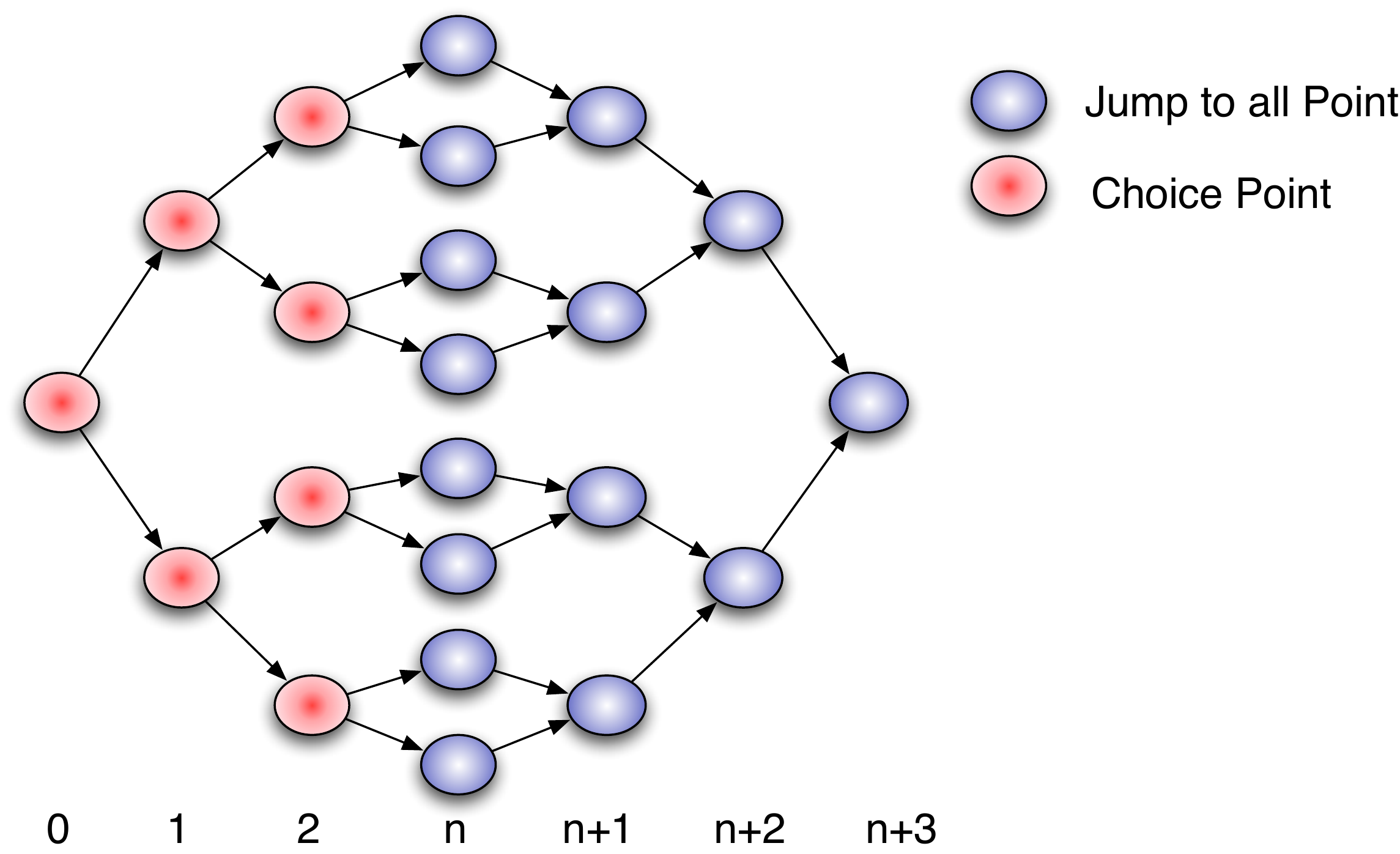}
}}
 \caption{A scalable-size score with $3.2^n - 2$ points.}
 \label{fig:Graph}
\end{figure}

We built automatically Interactive Scores (IS) with a number of points\footnote{The exact number of points is $3.2^n - 2$.} and relations in the order of $2^n$, with $n$ from two to ten (see fig. \ref{fig:Graph}). We ran each score 100 times as a stand-alone program. The duration of a time unit is determined by the time taken by Ntccrt to calculate the output, not by an external clock. The tests were performed on an iMac 2.6 GHz with 2 GB of RAM under Mac OS 10.5.7. It was compiled with GCC 4.2 and liked to Gecode 3.2.2.

\begin{figure}[!h]
 \centerline{\framebox{
 \includegraphics[width=\columnwidth]{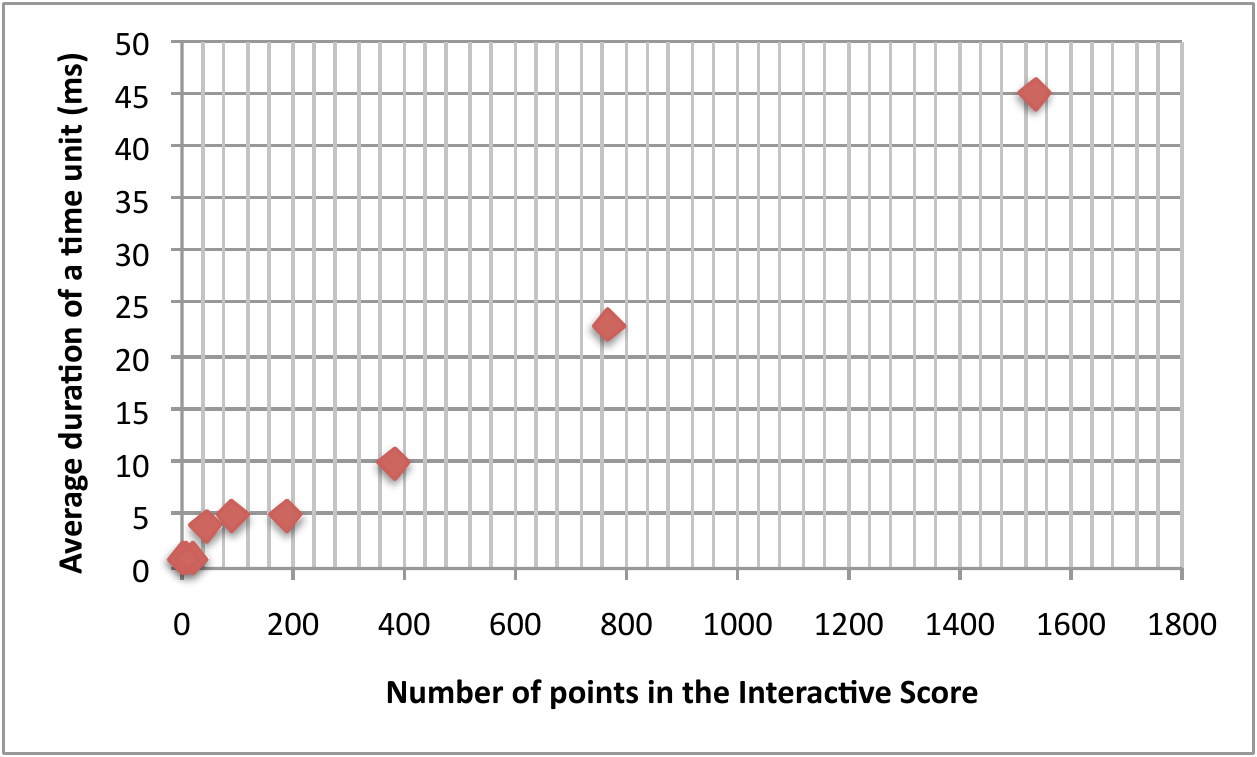}
}}
 \caption{Performance of the simulation fo the score in Fig. \ref{fig:Graph}.}
 \label{fig:Results}
\end{figure}

The authors of the Continuator \cite{continuator} argue that a multimedia interaction system with a response time less than 30 ms is able
to interact in real-time with even a very fast guitar jazz player.
Therefore, our results (fig. \ref{fig:Results}) are acceptable for real-time interaction with a guitarist for up to 1000 points (around 500 TOs).
We conjecture that a response time of 20 ms is appropriate to interact with a very fast percussionist. In that case, we can have up to 400 TOs.

\subsubsection{Usability of Ntccrt}
We found out intuitive to write \texttt{ntcc} models in Ntccrt, to someone familiar with \texttt{ntcc}, because it provides a Lisp interface with a syntax
similar to \texttt{ntcc}; 
for instance, \\$PredecessorWait$ is written as 

\begin{verbatim}
(defproc PredecessorsWait (i j) 
 (unlessp (v=? (ActivePoint i) j) 
  (||(call PredecessorsWait i j) 
    (tell= (ArrivedPoint j i) 1)))) 
\end{verbatim}

\noindent
It is slightly harder to write the same definition in C++

\begin{verbatim}
class predecessorsWait:public proc{ 
public:
AskBody* predecessorsWait::operator()(
Space* h, vector<int> intparameters, 
vector<variable *> variableparameters)
{return unless(eq(ActivePoint[i][j]), 
parallel(call(PredecessorWait,i,j),
tellEqual(ArrivedPoint[i][j],1)));}};
\end{verbatim}

\section{Concluding remarks}
We developed a model for multimedia interaction with con- ditional-branching and temporal relations based on
points and intervals. We implemented it using Ntccrt and Pure Data (Pd). We conclude from performance results that our prototype is compatible with real-time interaction for a reasonable amount of points and relations. An existing implementation of  Interactive Scores model is also capable of real-time and it can easy respect rigid durations, but 
such model does not support loops nor choice.

For simplicity, in our prototype we do not include hierarchy, we only model the interpretation $when$, we can only declare a single interval between two points, we can preserve rigid durations only in a few cases , and we can only execute a single instance
of an interval at the same time. 




An advantage of \texttt{ntcc} with respect to previous models of Interactive Scores, Pd, Max and Petri Nets is representing declarative conditions by the means of constraints.
Complex conditions, in particular those with an unknown number of parameters, are difficult to model in Max or Pd.
To model generic conditions in Max or Pd, we would have to define each condition either in a new patch or in a predefined library. In Petri nets, we would have to define a net for each condition. 

\subsection{Future work }
Ntccrt is not yet an interface for composers and designers of multimedia scenarios. For them is much more intuitive an interface such as Virage \cite{virage}. A graphical interface for our model should provide the means to specify the score as done in Example \ref{subsection:example}

Once we have the graphical interface, we plan to model audio processes in \texttt{ntcc} and replace them in the implementation by \textit{Faust} programs \cite{faust} which also have formal semantics. Using Faust, we can gain efficiency and preserve the formal properties of our model (see \cite{iclp2010} for a description of this idea).

\section{Acknowledgements}
We want to acknowledge Joseph Larralde and Camilo Rueda for their valuable comments on this model, and Andr\'{e}s Porras for his editing contributions.

\bibliographystyle{abbrv}


\end{document}